\documentclass[conference,10pt,twoside,twocolumn]{IEEEtran}

\usepackage[T1]{fontenc}
\usepackage{graphicx}
\usepackage{amssymb}
\usepackage{amsmath}
\usepackage{amsthm}
\usepackage{booktabs} 
\usepackage{multirow}
\usepackage{microtype}
\usepackage{cite}
\usepackage[caption=false]{subfig}
\usepackage{xcolor} 
\usepackage{url}
\usepackage{balance}
\usepackage{colortbl}
\usepackage[dvipsnames]{xcolor}

\usepackage{amssymb}
\usepackage{amsfonts}
\usepackage{mathrsfs}
\usepackage{xspace}
\usepackage{bm}
\usepackage{upgreek}

\newcommand{\safemath}[2]{\newcommand{#1}{\ensuremath{#2}\xspace}}

\safemath{\bma}{\mathbf{a}}
\safemath{\bmb}{\mathbf{b}}
\safemath{\bmc}{\mathbf{c}}
\safemath{\bmd}{\mathbf{d}}
\safemath{\bme}{\mathbf{e}}
\safemath{\bmf}{\mathbf{f}}
\safemath{\bmg}{\mathbf{g}}
\safemath{\bmh}{\mathbf{h}}
\safemath{\bmi}{\mathbf{i}}
\safemath{\bmj}{\mathbf{j}}
\safemath{\bmk}{\mathbf{k}}
\safemath{\bml}{\mathbf{l}}
\safemath{\bmm}{\mathbf{m}}
\safemath{\bmn}{\mathbf{n}}
\safemath{\bmo}{\mathbf{o}}
\safemath{\bmp}{\mathbf{p}}
\safemath{\bmq}{\mathbf{q}}
\safemath{\bmr}{\mathbf{r}}
\safemath{\bms}{\mathbf{s}}
\safemath{\bmt}{\mathbf{t}}
\safemath{\bmu}{\mathbf{u}}
\safemath{\bmv}{\mathbf{v}}
\safemath{\bmw}{\mathbf{w}}
\safemath{\bmx}{\mathbf{x}}
\safemath{\bmy}{\mathbf{y}}
\safemath{\bmz}{\mathbf{z}}
\safemath{\bmzero}{\mathbf{0}}
\safemath{\bmone}{\mathbf{1}}
\safemath{\Bell}{\ensuremath{\boldsymbol\ell}}

\bmdefine{\biad}{a}
\bmdefine{\bibd}{b}
\bmdefine{\bicd}{c}
\bmdefine{\bidd}{d}
\bmdefine{\bied}{e}
\bmdefine{\bifd}{f}
\bmdefine{\bigd}{g}
\bmdefine{\bihd}{h}
\bmdefine{\biid}{i}
\bmdefine{\bijd}{j}
\bmdefine{\bikd}{k}
\bmdefine{\bild}{l}
\bmdefine{\bimd}{m}
\bmdefine{\bind}{n}
\bmdefine{\biod}{o}
\bmdefine{\bipd}{p}
\bmdefine{\biqd}{q}
\bmdefine{\bird}{r}
\bmdefine{\bisd}{s}
\bmdefine{\bitd}{t}
\bmdefine{\biud}{u}
\bmdefine{\bivd}{v}
\bmdefine{\biwd}{w}
\bmdefine{\bixd}{x}
\bmdefine{\biyd}{y}
\bmdefine{\bizd}{z}

\bmdefine{\bixid}{\xi}
\bmdefine{\bilambdad}{\lambda}
\bmdefine{\bimud}{\mu}
\bmdefine{\bithetad}{\theta}
\bmdefine{\biphid}{\phi}
\bmdefine{\bideltad}{\delta}

\safemath{\bmia}{\biad}
\safemath{\bmib}{\bibd}
\safemath{\bmic}{\bicd}
\safemath{\bmid}{\bidd}
\safemath{\bmie}{\bied}
\safemath{\bmif}{\bifd}
\safemath{\bmig}{\bigd}
\safemath{\bmih}{\bihd}
\safemath{\bmii}{\biid}
\safemath{\bmij}{\bijd}
\safemath{\bmik}{\bikd}
\safemath{\bmil}{\bild}
\safemath{\bmim}{\bimd}
\safemath{\bmin}{\bind}
\safemath{\bmio}{\biod}
\safemath{\bmip}{\bipd}
\safemath{\bmiq}{\biqd}
\safemath{\bmir}{\bird}
\safemath{\bmis}{\bisd}
\safemath{\bmit}{\bitd}
\safemath{\bmiu}{\biud}
\safemath{\bmiv}{\bivd}
\safemath{\bmiw}{\biwd}
\safemath{\bmix}{\bixd}
\safemath{\bmiy}{\biyd}
\safemath{\bmiz}{\bizd}

\safemath{\bmxi}{\bixid}
\safemath{\bmlambda}{\bilambdad}
\safemath{\bmmu}{\bimud}
\safemath{\bmtheta}{\bithetad}
\safemath{\bmphi}{\biphid}
\safemath{\bmdelta}{\bideltad}

\safemath{\bA}{\mathbf{A}}
\safemath{\bB}{\mathbf{B}}
\safemath{\bC}{\mathbf{C}}
\safemath{\bD}{\mathbf{D}}
\safemath{\bE}{\mathbf{E}}
\safemath{\bF}{\mathbf{F}}
\safemath{\bG}{\mathbf{G}}
\safemath{\bH}{\mathbf{H}}
\safemath{\bI}{\mathbf{I}}
\safemath{\bJ}{\mathbf{J}}
\safemath{\bK}{\mathbf{K}}
\safemath{\bL}{\mathbf{L}}
\safemath{\bM}{\mathbf{M}}
\safemath{\bN}{\mathbf{N}}
\safemath{\bO}{\mathbf{O}}
\safemath{\bP}{\mathbf{P}}
\safemath{\bQ}{\mathbf{Q}}
\safemath{\bR}{\mathbf{R}}
\safemath{\bS}{\mathbf{S}}
\safemath{\bT}{\mathbf{T}}
\safemath{\bU}{\mathbf{U}}
\safemath{\bV}{\mathbf{V}}
\safemath{\bW}{\mathbf{W}}
\safemath{\bX}{\mathbf{X}}
\safemath{\bY}{\mathbf{Y}}
\safemath{\bZ}{\mathbf{Z}}

\safemath{\bZero}{\mathbf{0}}
\safemath{\bOne}{\mathbf{1}}
\safemath{\bDelta}{\mathbf{\Delta}}
\safemath{\bLambda}{\mathbf{\UpLambda}}
\safemath{\bPhi}{\mathbf{\Upphi}}
\safemath{\bSigma}{\mathbf{\Upsigma}}
\safemath{\bOmega}{\mathbf{\Upomega}}
\safemath{\bTheta}{\mathbf{\Uptheta}}

\bmdefine{\biAd}{A}
\bmdefine{\biBd}{B}
\bmdefine{\biCd}{C}
\bmdefine{\biDd}{D}
\bmdefine{\biEd}{E}
\bmdefine{\biFd}{F}
\bmdefine{\biGd}{G}
\bmdefine{\biHd}{H}
\bmdefine{\biId}{I}
\bmdefine{\biJd}{J}
\bmdefine{\biKd}{K}
\bmdefine{\biLd}{L}
\bmdefine{\biMd}{M}
\bmdefine{\biOd}{N}
\bmdefine{\biPd}{O}
\bmdefine{\biQd}{P}
\bmdefine{\biRd}{R}
\bmdefine{\biSd}{S}
\bmdefine{\biTd}{T}
\bmdefine{\biUd}{U}
\bmdefine{\biVd}{V}
\bmdefine{\biWd}{W}
\bmdefine{\biXd}{X}
\bmdefine{\biYd}{Y}
\bmdefine{\biZd}{Z}

\bmdefine{\biDelta}{\Delta}
\bmdefine{\biLambda}{\Lambda}
\bmdefine{\biPhi}{\Phi}
\bmdefine{\biSigma}{\Sigma}
\bmdefine{\biOmega}{\Omega}
\bmdefine{\biTheta}{\Theta}

\safemath{\bimA}{\biAd}
\safemath{\bimB}{\biBd}
\safemath{\bimC}{\biCd}
\safemath{\bimD}{\biDd}
\safemath{\bimE}{\biEd}
\safemath{\bimF}{\biFd}
\safemath{\bimG}{\biGd}
\safemath{\bimH}{\biHd}
\safemath{\bimI}{\biId}
\safemath{\bimJ}{\biJd}
\safemath{\bimK}{\biKd}
\safemath{\bimL}{\biLd}
\safemath{\bimM}{\biMd}
\safemath{\bimN}{\biNd}
\safemath{\bimO}{\biOd}
\safemath{\bimP}{\biPd}
\safemath{\bimQ}{\biQd}
\safemath{\bimR}{\biRd}
\safemath{\bimS}{\biSd}
\safemath{\bimT}{\biTd}
\safemath{\bimU}{\biUd}
\safemath{\bimV}{\biVd}
\safemath{\bimW}{\biWd}
\safemath{\bimX}{\biXd}
\safemath{\bimY}{\biYd}
\safemath{\bimZ}{\biZd}

\safemath{\bimDelta}{\biDelta}
\safemath{\bimLambda}{\biLambda}
\safemath{\bimPhi}{\biPhi}
\safemath{\bimSigma}{\biSigma}
\safemath{\bimOmega}{\biOmega}
\safemath{\bimTheta}{\biTheta}

\safemath{\setA}{\mathcal{A}}
\safemath{\setB}{\mathcal{B}}
\safemath{\setC}{\mathcal{C}}
\safemath{\setD}{\mathcal{D}}
\safemath{\setE}{\mathcal{E}}
\safemath{\setF}{\mathcal{F}}
\safemath{\setG}{\mathcal{G}}
\safemath{\setH}{\mathcal{H}}
\safemath{\setI}{\mathcal{I}}
\safemath{\setJ}{\mathcal{J}}
\safemath{\setK}{\mathcal{K}}
\safemath{\setL}{\mathcal{L}}
\safemath{\setM}{\mathcal{M}}
\safemath{\setN}{\mathcal{N}}
\safemath{\setO}{\mathcal{O}}
\safemath{\setP}{\mathcal{P}}
\safemath{\setQ}{\mathcal{Q}}
\safemath{\setR}{\mathcal{R}}
\safemath{\setS}{\mathcal{S}}
\safemath{\setT}{\mathcal{T}}
\safemath{\setU}{\mathcal{U}}
\safemath{\setV}{\mathcal{V}}
\safemath{\setW}{\mathcal{W}}
\safemath{\setX}{\mathcal{X}}
\safemath{\setY}{\mathcal{Y}}
\safemath{\setZ}{\mathcal{Z}}
\safemath{\emptySet}{\varnothing}

\safemath{\colA}{\mathscr{A}}
\safemath{\colB}{\mathscr{B}}
\safemath{\colC}{\mathscr{C}}
\safemath{\colD}{\mathscr{D}}
\safemath{\colE}{\mathscr{E}}
\safemath{\colF}{\mathscr{F}}
\safemath{\colG}{\mathscr{G}}
\safemath{\colH}{\mathscr{H}}
\safemath{\colI}{\mathscr{I}}
\safemath{\colJ}{\mathscr{J}}
\safemath{\colK}{\mathscr{K}}
\safemath{\colL}{\mathscr{L}}
\safemath{\colM}{\mathscr{M}}
\safemath{\colN}{\mathscr{N}}
\safemath{\colO}{\mathscr{O}}
\safemath{\colP}{\mathscr{P}}
\safemath{\colQ}{\mathscr{Q}}
\safemath{\colR}{\mathscr{R}}
\safemath{\colS}{\mathscr{S}}
\safemath{\colT}{\mathscr{T}}
\safemath{\colU}{\mathscr{U}}
\safemath{\colV}{\mathscr{V}}
\safemath{\colW}{\mathscr{W}}
\safemath{\colX}{\mathscr{X}}
\safemath{\colY}{\mathscr{Y}}
\safemath{\colZ}{\mathscr{Z}}

\safemath{\opA}{\mathbb{A}}
\safemath{\opB}{\mathbb{B}}
\safemath{\opC}{\mathbb{C}}
\safemath{\opD}{\mathbb{D}}
\safemath{\opE}{\mathbb{E}}
\safemath{\opF}{\mathbb{F}}
\safemath{\opG}{\mathbb{G}}
\safemath{\opH}{\mathbb{H}}
\safemath{\opI}{\mathbb{I}}
\safemath{\opJ}{\mathbb{J}}
\safemath{\opK}{\mathbb{K}}
\safemath{\opL}{\mathbb{L}}
\safemath{\opM}{\mathbb{M}}
\safemath{\opN}{\mathbb{N}}
\safemath{\opO}{\mathbb{O}}
\safemath{\opP}{\mathbb{P}}
\safemath{\opQ}{\mathbb{Q}}
\safemath{\opR}{\mathbb{R}}
\safemath{\opS}{\mathbb{S}}
\safemath{\opT}{\mathbb{T}}
\safemath{\opU}{\mathbb{U}}
\safemath{\opV}{\mathbb{V}}
\safemath{\opW}{\mathbb{W}}
\safemath{\opX}{\mathbb{X}}
\safemath{\opY}{\mathbb{Y}}
\safemath{\opZ}{\mathbb{Z}}
\safemath{\opZero}{\mathbb{O}}
\safemath{\identityop}{\opI}

\safemath{\veca}{\bma}
\safemath{\vecb}{\bmb}
\safemath{\vecc}{\bmc}
\safemath{\vecd}{\bmd}
\safemath{\vece}{\bme}
\safemath{\vecf}{\bmf}
\safemath{\vecg}{\bmg}
\safemath{\vech}{\bmh}
\safemath{\veci}{\bmi}
\safemath{\vecj}{\bmj}
\safemath{\veck}{\bmk}
\safemath{\vecl}{\bml}
\safemath{\vecm}{\bmm}
\safemath{\vecn}{\bmn}
\safemath{\veco}{\bmo}
\safemath{\vecp}{\bmp}
\safemath{\vecq}{\bmq}
\safemath{\vecr}{\bmr}
\safemath{\vecs}{\bms}
\safemath{\vect}{\bmt}
\safemath{\vecu}{\bmu}
\safemath{\vecv}{\bmv}
\safemath{\vecw}{\bmw}
\safemath{\vecx}{\bmx}
\safemath{\vecy}{\bmy}
\safemath{\vecz}{\bmz}

\safemath{\veczero}{\bmzero}
\safemath{\vecone}{\bmone}
\safemath{\vecxi}{\bmxi}
\safemath{\veclambda}{\bmlambda}
\safemath{\vecmu}{\bmmu}
\safemath{\vectheta}{\bmtheta}
\safemath{\vecphi}{\bmphi}
\safemath{\vecdelta}{\bmdelta}

\safemath{\matA}{\bA}
\safemath{\matB}{\bB}
\safemath{\matC}{\bC}
\safemath{\matD}{\bD}
\safemath{\matE}{\bE}
\safemath{\matF}{\bF}
\safemath{\matG}{\bG}
\safemath{\matH}{\bH}
\safemath{\matI}{\bI}
\safemath{\matJ}{\bJ}
\safemath{\matK}{\bK}
\safemath{\matL}{\bL}
\safemath{\matM}{\bM}
\safemath{\matN}{\bN}
\safemath{\matO}{\bO}
\safemath{\matP}{\bP}
\safemath{\matQ}{\bQ}
\safemath{\matR}{\bR}
\safemath{\matS}{\bS}
\safemath{\matT}{\bT}
\safemath{\matU}{\bU}
\safemath{\matV}{\bV}
\safemath{\matW}{\bW}
\safemath{\matX}{\bX}
\safemath{\matY}{\bY}
\safemath{\matZ}{\bZ}
\safemath{\matzero}{\bmzero}

\safemath{\matDelta}{\bDelta}
\safemath{\matLambda}{\bLambda}
\safemath{\matPhi}{\bPhi}
\safemath{\matSigma}{\bSigma}
\safemath{\matOmega}{\bOmega}
\safemath{\matTheta}{\bTheta}

\safemath{\matidentity}{\matI}
\safemath{\matone}{\matO}

\safemath{\rnda}{A}
\safemath{\rndb}{B}
\safemath{\rndc}{C}
\safemath{\rndd}{D}
\safemath{\rnde}{E}
\safemath{\rndf}{F}
\safemath{\rndg}{G}
\safemath{\rndh}{H}
\safemath{\rndi}{I}
\safemath{\rndj}{J}
\safemath{\rndk}{K}
\safemath{\rndl}{L}
\safemath{\rndm}{M}
\safemath{\rndn}{N}
\safemath{\rndo}{O}
\safemath{\rndp}{P}
\safemath{\rndq}{Q}
\safemath{\rndr}{R}
\safemath{\rnds}{S}
\safemath{\rndt}{T}
\safemath{\rndu}{U}
\safemath{\rndv}{V}
\safemath{\rndw}{W}
\safemath{\rndx}{X}
\safemath{\rndy}{Y}
\safemath{\rndz}{Z}

\safemath{\rveca}{\bimA}
\safemath{\rvecb}{\bimB}
\safemath{\rvecc}{\bimC}
\safemath{\rvecd}{\bimD}
\safemath{\rvece}{\bimE}
\safemath{\rvecf}{\bimF}
\safemath{\rvecg}{\bimG}
\safemath{\rvech}{\bimH}
\safemath{\rveci}{\bimI}
\safemath{\rvecj}{\bimJ}
\safemath{\rveck}{\bimK}
\safemath{\rvecl}{\bimL}
\safemath{\rvecm}{\bimM}
\safemath{\rvecn}{\bimN}
\safemath{\rveco}{\bomO}
\safemath{\rvecp}{\bimP}
\safemath{\rvecq}{\bimQ}
\safemath{\rvecr}{\bimR}
\safemath{\rvecs}{\bimS}
\safemath{\rvect}{\bimT}
\safemath{\rvecu}{\bimU}
\safemath{\rvecv}{\bimV}
\safemath{\rvecw}{\bimW}
\safemath{\rvecx}{\bimX}
\safemath{\rvecy}{\bimY}
\safemath{\rvecz}{\bimZ}

\safemath{\rvecxi}{\bmxi}
\safemath{\rveclambda}{\bmlambda}
\safemath{\rvecmu}{\bmmu}
\safemath{\rvectheta}{\bmtheta}
\safemath{\rvecphi}{\bmphi}

\safemath{\rmatA}{\bimA}
\safemath{\rmatB}{\bimB}
\safemath{\rmatC}{\bimC}
\safemath{\rmatD}{\bimD}
\safemath{\rmatE}{\bimE}
\safemath{\rmatF}{\bimF}
\safemath{\rmatG}{\bimG}
\safemath{\rmatH}{\bimH}
\safemath{\rmatI}{\bimI}
\safemath{\rmatJ}{\bimJ}
\safemath{\rmatK}{\bimK}
\safemath{\rmatL}{\bimL}
\safemath{\rmatM}{\bimM}
\safemath{\rmatN}{\bimN}
\safemath{\rmatO}{\bimO}
\safemath{\rmatP}{\bimP}
\safemath{\rmatQ}{\bimQ}
\safemath{\rmatR}{\bimR}
\safemath{\rmatS}{\bimS}
\safemath{\rmatT}{\bimT}
\safemath{\rmatU}{\bimU}
\safemath{\rmatV}{\bimV}
\safemath{\rmatW}{\bimW}
\safemath{\rmatX}{\bimX}
\safemath{\rmatY}{\bimY}
\safemath{\rmatZ}{\bimZ}

\safemath{\rmatDelta}{\bimDelta}
\safemath{\rmatLambda}{\bimLambda}
\safemath{\rmatPhi}{\bimPhi}
\safemath{\rmatSigma}{\bimSigma}
\safemath{\rmatOmega}{\bimOmega}
\safemath{\rmatTheta}{\bimTheta}

\usepackage{amssymb}
\usepackage{amsfonts}
\usepackage{mathrsfs}
\usepackage{xspace}
\usepackage{bm}
\usepackage{fancyref}
\usepackage{textcomp}

\usepackage{multirow}
\usepackage{stmaryrd}

\newenvironment{textbmatrix}{	\setlength{\arraycolsep}{2.5pt}%
								\left[\begin{matrix}}{\end{matrix}\right]%
								\raisebox{0.08ex}{\vphantom{M}}}

\def\be{\begin{equation}}
\def\ee{\end{equation}}
\def\een{\nonumber \end{equation}}
\def\mat{\begin{bmatrix}}
\def\emat{\end{bmatrix}}
\def\btm{\begin{textbmatrix}}
\def\etm{\end{textbmatrix}}

\def\ba#1\ea{\begin{align}#1\end{align}}
\def\bas#1\eas{\begin{align*}#1\end{align*}}
\def\bs#1\es{\begin{split}#1\end{split}}
\def\bg#1\eg{\begin{gather}#1\end{gather}}
\def\bml#1\eml{\begin{multline}#1\end{multline}}
\def\bi#1\ei{\begin{itemize}#1\end{itemize}}

\newcommand{\lefto}{\mathopen{}\left}

\DeclareMathOperator{\Exop}{\opE}			

\DeclareMathOperator{\erf}{erf}				

\newcommand{\Ex}[1]{\ensuremath{\Exop\lefto[#1\right]}} 	

\safemath{\dirac}{\delta}					
\safemath{\krond}{\dirac}					

\safemath{\upto}{\uparrow}
\safemath{\downto}{\downarrow}
\safemath{\iu}{j}							
\safemath{\ev}{\lambda}						
\safemath{\hilseqspace}{l^{2}}				
\newcommand{\banachfunspace}[1]{\setL^{#1}}	
\safemath{\hilfunspace}{\banachfunspace{2}}	

\safemath{\SNR}{\textit{SNR}} 				
\safemath{\PAR}{\textit{PAR}} 				
\safemath{\No}{N_0}							
\safemath{\Es}{E_s}							
\safemath{\Eb}{E_b}							
\safemath{\EbNo}{\frac{\Eb}{\No}}
\safemath{\EsNo}{\frac{\Es}{\No}}

\DeclareMathOperator{\CHop}{\ensuremath{\opH}} 
\safemath{\tvir}{\rndh_{\CHop}}				
\safemath{\tvtf}{\rndl_{\CHop}}				
\safemath{\spf}{\rnds_{\CHop}}				
\safemath{\bff}{H_{\CHop}}					

\safemath{\ircf}{r_{h}}						
\safemath{\tftvcf}{r_{s}}					
\safemath{\tfcf}{r_{l}}						
\safemath{\bfcf}{r_{H}}						

\safemath{\tcorr}{c_h}						
\safemath{\scf}{c_{s}}						
\safemath{\tfcorr}{c_{l}}					
\safemath{\fcorr}{c_{H}}						

\safemath{\mi}{I}							
\safemath{\capacity}{C}						

\safemath{\normal}{\mathcal{N}}			
\safemath{\jpg}{\mathcal{CN}}			
\safemath{\mchain}{\leftrightarrow}		

\safemath{\dB}{\,\mathrm{dB}}
\safemath{\dBm}{\,\mathrm{dBm}}
\safemath{\Hz}{\,\mathrm{Hz}}
\safemath{\kHz}{\,\mathrm{kHz}}
\safemath{\MHz}{\,\mathrm{MHz}}
\safemath{\GHz}{\,\mathrm{GHz}}
\safemath{\s}{\,\mathrm{s}}
\safemath{\ms}{\,\mathrm{ms}}
\safemath{\mus}{\,\mathrm{\text{\textmu}s}}
\safemath{\ns}{\,\mathrm{ns}}
\safemath{\ps}{\,\mathrm{ps}}
\safemath{\meter}{\,\mathrm{m}}
\safemath{\mm}{\,\mathrm{mm}}
\safemath{\cm}{\,\mathrm{cm}}
\safemath{\m}{\,\mathrm{m}}
\safemath{\W}{\,\mathrm{W}}
\safemath{\mW}{\, \mathrm{mW}}
\safemath{\J}{\,\mathrm{J}}
\safemath{\K}{\,\mathrm{K}}
\safemath{\bit}{\,\mathrm{bit}}
\safemath{\nat}{\,\mathrm{nat}}

\safemath{\define}{\triangleq}			

\safemath{\equivalent}{\sim}
\safemath{\distas}{\sim}					
\safemath{\sdiff}{\Delta}				

\safemath{\reals}{\mathbb{R}}
\safemath{\positivereals}{\reals_{+}}
\safemath{\integers}{\mathbb{Z}}
\safemath{\posint}{\integers_{+}}
\safemath{\naturals}{\mathbb{N}}
\safemath{\posnaturals}{\naturals_{+}}
\safemath{\complexset}{\mathbb{C}}
\safemath{\rationals}{\mathbb{Q}}

\newcommand*{\fancyrefapplabelprefix}{app}		
\newcommand*{\fancyrefthmlabelprefix}{thm}		
\newcommand*{\fancyreflemlabelprefix}{lem}		
\newcommand*{\fancyrefcorlabelprefix}{cor}		
\newcommand*{\fancyrefdeflabelprefix}{def}		
\newcommand*{\fancyrefproplabelprefix}{prop}		
\newcommand*{\fancyrefexmpllabelprefix}{exmpl}
\newcommand*{\fancyrefalglabelprefix}{alg}		
\newcommand*{\fancyreftbllabelprefix}{tbl}		

\frefformat{vario}{\fancyrefseclabelprefix}{Sec.~#1}
\frefformat{vario}{\fancyrefthmlabelprefix}{Theorem~#1}
\frefformat{vario}{\fancyreftbllabelprefix}{Tbl.~#1}
\frefformat{vario}{\fancyreflemlabelprefix}{Lemma~#1}
\frefformat{vario}{\fancyrefcorlabelprefix}{Corollary~#1}
\frefformat{vario}{\fancyrefdeflabelprefix}{Definition~#1}
\frefformat{vario}{\fancyreffiglabelprefix}{Fig.~#1}
\frefformat{vario}{\fancyrefapplabelprefix}{Appendix~#1}
\frefformat{vario}{\fancyrefeqlabelprefix}{(#1)}
\frefformat{vario}{\fancyrefproplabelprefix}{Proposition~#1}
\frefformat{vario}{\fancyrefexmpllabelprefix}{Example~#1}
\frefformat{vario}{\fancyrefalglabelprefix}{Algorithm~#1}

 \newtheorem{thm}{Theorem}

 \newtheorem{lem}[thm]{Lemma}

\safemath{\dictab}{[\,\dicta\,\,\dictb\,]}

\safemath{\ysig}{\bmy}
\safemath{\ysighat}{\hat{\ysig}}
\safemath{\ysigdim}{M}
\safemath{\xsig}{\bmx}
\safemath{\xsigdim}{N}
\safemath{\nx}{n_x}
\safemath{\zsig}{\bmz}
\safemath{\zsigdim}{\ysigdim}
\safemath{\rsig}{\bmr}
\safemath{\Adict}{\bA}
\safemath{\Adicttilde}{\widetilde{\Adict}}
\safemath{\Adictdim}{\outputdim\times\xsigdim}
\safemath{\avec}{\bma}
\safemath{\avectilde}{\tilde{\avec}}
\safemath{\Bdict}{\bB}
\safemath{\Bdicttilde}{\widetilde{\Bdict}}
\safemath{\Cdict}{\bC}
\safemath{\cvec}{\bmc}
\safemath{\Ddict}{\bD}
\safemath{\Ddictdim}{\ysigdim\times\xsigdim}
\safemath{\dvec}{\bmd}
\safemath{\Ddicttilde}{\widetilde{\bD}}
\safemath{\Bonb}{\bB}
\safemath{\bvec}{\bmb}
\safemath{\Bonbdim}{\ysigdim\times\ysigdim}
\safemath{\noise}{\bmn}
\safemath{\noisedim}{\ysigim}
\safemath{\err}{\bme}
\safemath{\errdim}{\ysigdim}
\safemath{\errset}{\setE}
\safemath{\nerr}{n_e}
\safemath{\delop}{\bP_\errset}
\safemath{\delopc}{\bP_{{\errset}^c}}

\safemath{\cplxi}{\imath}
\safemath{\cplxj}{\jmath}

\safemath{\dict}{\matD}
\safemath{\inputdim}{N}		
\safemath{\outputdim}{M}		
\safemath{\sparsity}{S}	
\safemath{\inputdimA}{{N_a}}	
\safemath{\inputdimB}{{N_b}}	
\safemath{\elemA}{{n_a}}	
\safemath{\elemB}{{n_b}}	
\safemath{\resA}{\matR_a}	
\safemath{\resB}{\matR_b}	
\safemath{\subD}{\matS} 
\safemath{\subA}{\matS_a} 
\safemath{\subB}{\matS_b} 
\safemath{\dicta}{\matA} 	
\safemath{\dictb}{\matB} 	
\safemath{\hollowS}{H}
\safemath{\hollowA}{H_a}
\safemath{\hollowB}{H_b}
\safemath{\cross}{Z}
\safemath{\coh}{\mu_d}			
\safemath{\coha}{\mu_a}			
\safemath{\cohb}{\mu_b}			
\safemath{\mubs}{\nu}	
\safemath{\cohm}{\mu_m} 
\safemath{\dictset}{\setD}	
\safemath{\dictsetp}{\dictset(\coh,\coha,\cohb)}	
\safemath{\dictsetgen}{\dictset_\text{gen}}
\safemath{\dictsetgenp}{\dictsetgen(\coh)}
\safemath{\dictsetonb}{\dictset_\text{onb}}
\safemath{\dictsetonbp}{\dictsetonb(\coh)}

\safemath{\leftside}{U}
\safemath{\rightsideA}{R_a}
\safemath{\rightsideB}{R_b}

\safemath{\indexS}{\setI_S} 

\safemath{\na}{n_a}			
\safemath{\nb}{n_b}			
\safemath{\coeffa}{p_i}	
\safemath{\coeffb}{q_j}	
\safemath{\seta}{\setP}		
\safemath{\setb}{\setQ}     
\safemath{\setw}{\setW}	
\safemath{\setz}{\setZ}	
\safemath{\cola}{\veca}		
\safemath{\colb}{\vecb}		
\safemath{\cold}{\vecd}		
\safemath{\inputvec}{\vecx} 	
\safemath{\error}{\vece}	
\safemath{\noiseout}{\vecz} 	
\safemath{\inputvecel}{x}
\safemath{\inputveca}{\vecx_a}
\safemath{\inputvecb}{\vecx_b}
\safemath{\outputvec}{\vecy}	
\safemath{\lambdamin}{\lambda_{\mathrm{min}}}

\safemath{\elltwo}{\ell_2}
\safemath{\ellone}{\ell_1}
\safemath{\ellzero}{\ell_0}
\safemath{\ellinf}{\ell_\infty}
\safemath{\ellinftilde}{\ell_{\widetilde\infty}}
\safemath{\licard}{Z(\coh,\coha,\cohb)}
\safemath{\xsol}{\hat{x}}
\safemath{\xbord}{x_b}		
\safemath{\xstat}{x_s}		
\safemath{\xstatLone}{\tilde{x}_s}
\safemath{\order}{\mathcal{O}} 
\safemath{\scales}{\Theta} 
\safemath{\ones}{\mathbf{1}} 
\safemath{\zeroes}{\mathbf{0}} 
\safemath{\thlone}{\kappa(\coh,\cohb)} 
\safemath{\constoneA}{\delta} 
\safemath{\constoneB}{\epsilon} 
\safemath{\nlarge}{L}				   
\safemath{\sumlarge}{S_\nlarge}
\safemath{\maxlarger}{P_\nlarge}	   
\safemath{\Pzero}{\textrm{P0}}	
\safemath{\Pone}{\textrm{P1}}
\safemath{\vecfir}{\vecw}			 
\safemath{\vecsec}{\vecz}
\safemath{\elvecfir}{w}              
\safemath{\elvecsec}{z}				 
\safemath{\nlargefir}{n}
\safemath{\normout}{\gamma}
\safemath{\auxfun}{h}
\safemath{\supp}{\textrm{supp}}

\safemath{\indexa}{\ell}
\safemath{\indexb}{r}
\safemath{\indexc}{i}
\safemath{\indexd}{j}

\safemath{\project}{P}

\usepackage{framed}

\IEEEoverridecommandlockouts

\newcommand{\fs}{f_\textnormal{s}}
\newcommand{\fmax}{f_\textnormal{max}}
\newcommand{\fsig}{f_\textnormal{sig}}
\newcommand{\eexp}{\mathrm{e}}
\newcommand{\jmod}{\;\mathrm{mod}\;}
\newcommand{\nth}{$n\mkern2mu$th\xspace}
\safemath{\dBFS}{\,\mathrm{dBFS}}
\safemath{\dBc}{\,\mathrm{dBc}}
\newcommand{\modulo}{\mathbin{\texttt{\%}}}
\begin{document}

\title{Spectral Impact of Mismatches in Interleaved ADCs}

\author{J\'er\'emy Guichemerre$^*$, Robert Reutemann$^\diamondsuit$, Thomas Burger$^*$, and Christoph Studer$^*$\\[0.1cm]
\textit{$^*$ETH Zurich, Switzerland; $^\diamondsuit$Miromico IC AG, Zurich, Switzerland} \\
\textit{emails: jeremyg@iis.ee.ethz.ch, rre@miromico-ic.ch, burgert@ethz.ch, studer@ethz.ch}
}

\maketitle


\begin{abstract}
Interleaved ADCs are critical for applications requiring multi-gigasample per second (GS/s) rates, but their performance is often limited by offset, gain, and timing skew mismatches across the sub-ADCs.
We propose exact but compact expressions that describe the impact of each of those non-idealities on the output spectrum.
We derive the distribution of the power of the induced spurs and replicas, critical for yield-oriented derivation of sub-ADC specifications.
Finally, we provide a practical example in which calibration step sizes are derived under the constraint of a target production yield.
\end{abstract}

\section{Introduction}

Time interleaving enables analog-to-digital converters (ADCs) to reach multi-gigasample per second (GS/s) rates~\cite{Black_1980,adc_survey}.
Such high sampling rates are essential in applications such as high-speed wireline transceivers, direct RF sampling in wireless receivers, and radar systems.
The drawback of interleaving is that mismatches among the sub-ADCs introduce distortion: affine errors at the sub-ADC level produce spurious tones and replicas at the system level.
The most critical mismatches in real-world sub-ADC designs are offset, gain, and timing skew.
Offset corresponds to a constant shift at the sub-ADC's output, creating spurious tones at multiples of the sub-ADC rate.
Gain mismatches are scaling errors and yield replicas of the input spectrum.
Skews are sampling-time errors and replicate a differentiated version of the input spectrum.
As those mismatches are fixed after fabrication (or vary slowly), performance metrics and their design targets must be linked to production yield---knowledge of the distribution of the impact mismatches have on the output spectrum is therefore necessary.

\subsection{Contributions}

In this paper, we express the impact of mismatches through the discrete Fourier transform (DFT) of the mismatch sequences to derive compact but rigorous expressions that capture the effects of offset, gain, and skew mismatches without relying on tone-based approximations.
This provides simple and intuitive spectral expressions, which naturally extend to a statistical framework.
We derive exact spur and replica power distributions under Gaussian mismatch assumptions, and further quantify the accuracy of Gaussian approximations in the case of uniform distributions for practical interleaving factors.

\subsection{Limitations of Prior Art}

Vogel~\cite{Vogel_2005} derived expressions for the average SNDR, but expectations alone prevent a yield analysis.
Ghosh~\cite{Ghosh_2021} provided an exact derivation of SNDR degradation, but the method is intricate and restricted to single-tone inputs.
Neither approach provides the statistical framework needed for  yield-driven design.
Monte--Carlo analysis alone is also insufficient, as system-level exploration requires fast and general predictions.

\subsection{Notation} \label{sec:1_notation}

We write matrices in bold uppercase and column vectors in bold lowercase.
We index the entries $u_k$ of a vector $\bmu\in\complexset^N$ with $k$ ranging from $0$ to $N\!-\!1$, and we write $k \modulo N$ for $k \jmod N$.
The $N\!\times\!N$ identity matrix is $\bI_N$.
We define the discrete Fourier transform (DFT) and the DFT matrix~$\bF$ as
\begin{align} \label{eq:1_DFT}
    \tilde{u}_k = \{\bF\bmu\}_k
        = \textstyle \frac{1}{N}\sum_{n=0}^{N-1} \eexp^{-2\pi j \frac{kn}{N}}u_n,
\end{align}
where $\tilde{\bmu}\in\complexset^N$ is the DFT of $\bmu\in\complexset^N$ and $j$ is the imaginary unit.
The $1/N$ normalization in \fref{eq:1_DFT} ensures that the squared magnitudes $|\tilde{u}_k|^2$ represent average powers, independent of the length $N$ of the sequence.
We denote the Dirac distribution as~$\delta(t)$ for $t\in\reals$; the convolution of $g(t)$ with $h(t)$ is $g(t) \ast h(t)$.
We normalize the full-scale range of ADCs to $[-1,1]$;  $0\dBFS$ is the power of a full-scale sine wave. Zero-mean real and complex circularly-symmetric Gaussian distributions of variance $\sigma^2$ are $\normal(0, \sigma^2)$ and $\jpg(0, \sigma^2)$, respectively. The Gaussian error function is denoted by~$\erf(x)$.


\section{Ideal Interleaving and Mismatches} \label{sec:2}

We first analyze the Fourier-domain representation of an ideal interleaved ADC, showing how the outputs of the sub-ADCs perfectly recombine (see~\fref{fig:2_block}).
We then extend our analysis to consider offset, gain, and skew mismatches, and derive expressions that characterize their output spectrum impact.

\subsection{Ideal Interleaving} \label{sec:2_ideal}

\begin{figure*}[tp]
    \centering

    \begin{minipage}[c]{0.32\textwidth}
        \centering
        \subfloat[]{%
            \includegraphics[width=\linewidth]{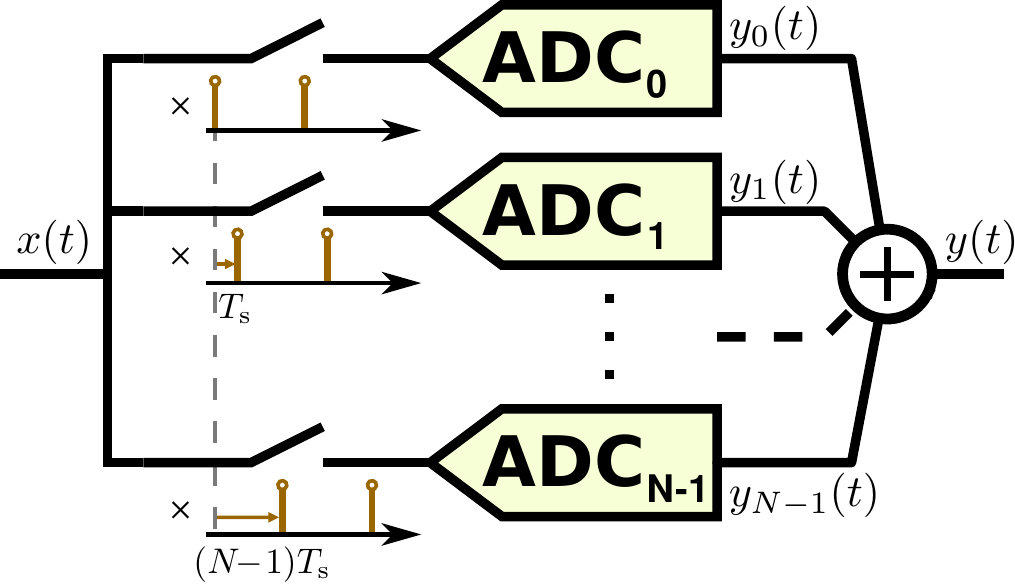}%
            \label{fig:2_block}}
    \end{minipage}
    \hfill
    \begin{minipage}[c]{0.32\textwidth}
        \centering
        \subfloat[]{%
            \includegraphics[width=\linewidth]{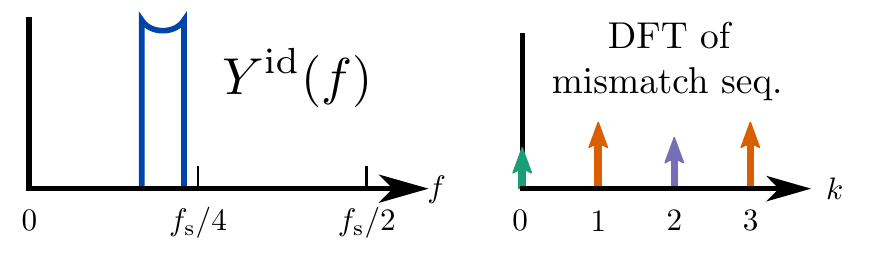}%
            \vspace{-0.15cm}
            \label{fig:2_ideal_DFT}}%
        \vspace{-0.1cm} 
        \\[0.3em]
        \subfloat[]{%
            \includegraphics[width=\linewidth]{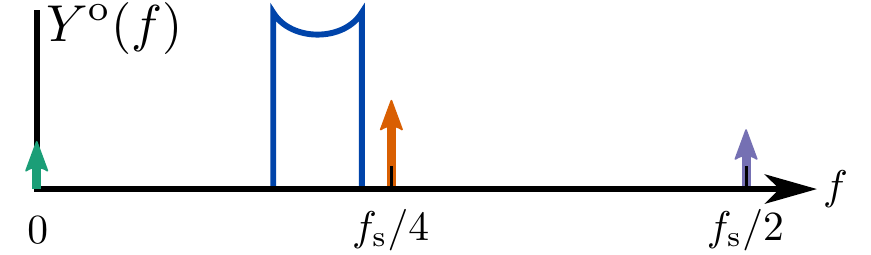}%
            \label{fig:2_offset}}%
        \vspace{-0.1cm} 
    \end{minipage}
    \hfill
    \begin{minipage}[c]{0.32\textwidth}
        \centering
        \subfloat[]{%
            \includegraphics[width=\linewidth]{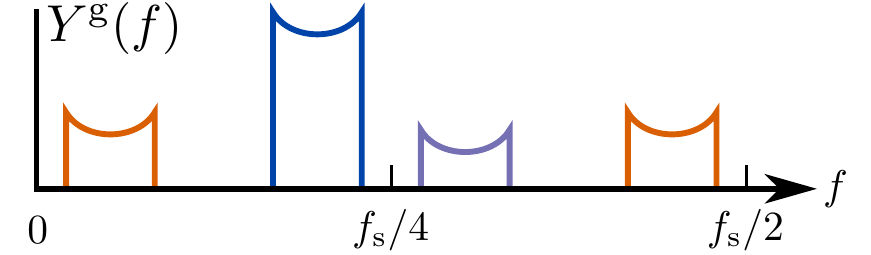}%
            \label{fig:2_gain}}%
        \vspace{-0.1cm} 
        \\[0.3em]
        \subfloat[]{%
            \includegraphics[width=\linewidth]{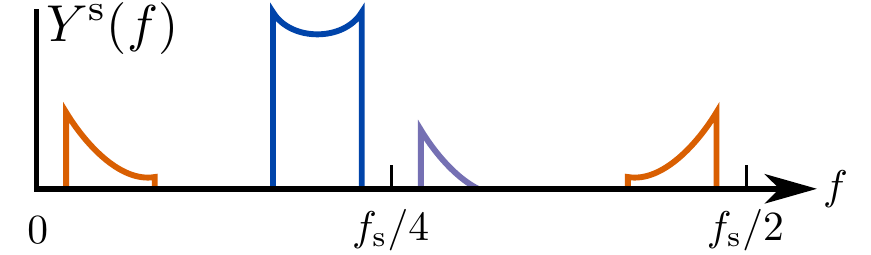}%
            \label{fig:2_skew}}%
        \vspace{-0.1cm} 
    \end{minipage}

    \vspace{-0.05cm}
    \caption{(a) Conceptual block-diagram of an interleaved ADC, (b) example of an ideally sampled single-sided output spectrum and DFT of a mismatch sequence for a $4\times$ interleaved ADC. (c), (d), and (e) show the impact of the mismatch sequence of (b) in the case of offset, gain mismatch, and timing skew, respectively.}
    \label{fig:2}
\end{figure*}

We consider an interleaved ADC consisting of $N$ sub-ADCs, each operating at a sampling rate of $\fs/N$. When sampling an input $x(t)$, the sampled output of the ideal \nth sub-ADC is
\begin{align} \label{eq:2_x}
    \textstyle y_n^\textnormal{id}(t)= x(t) \sum_k \delta \!\left( t-k\frac{N}{\fs}-\frac{n}{\fs} \right)\!.
\end{align}
Let $X(f)$ be the Fourier transform (FT) of $x(t)$. Then, the FT of the  \nth  sub-ADC output $y_n^\textnormal{id}$ is given by
\begin{align} \label{eq:2_X}
    \textstyle Y_n^\textnormal{id}\!\left(f\right) = 
    X\!\left(f\right) \ast \frac{\fs}{N} \sum_k \mathrm{e}^{-j2\pi \frac{kn}{N}} \delta \!\left( f-k\frac{\fs}{N} \right),
\end{align}
where the convolution with the Dirac comb leads to the expected aliasing corresponding to the $\fs/N$ rate.
However, when summing the output of all $N$ sub-ADCs, the resulting FT should be identical to that of an $\fs$-rate ADC.
Indeed, from~\fref{eq:2_X}, we obtain
\begin{align}
    Y^\textnormal{id}(f) = & \textstyle \sum_{n=0}^{N-1} Y_n^\textnormal{id}\!\left(f\right) \nonumber \\
         =\, & \textstyle X\!\left(f\right) \ast \frac{\fs}{N} \sum_k \!\left[ \delta \!\left( f-k\frac{\fs}{N} \right) \sum_{n=0}^{N-1} \mathrm{e}^{-j2\pi \frac{kn}{N}}\right]\!. \label{eq:2_sumX}
\end{align}
The sum in $n$ in the right-hand side corresponds to the sum of the $N\mkern2mu$th roots of unity.
Therefore, when $k$ is a multiple of $N$, the sum equals $N$; in any other case, the sum equals zero.
We re-index the sum to ignore the zero entries and obtain
\begin{align}
    \textstyle Y^\textnormal{id}(f) = X\!\left(f\right) \ast \fs \sum_k \delta \!\left( f-k\fs \right),
\end{align}
which matches the output of an ideal ADC of rate $\fs$; see~\fref{fig:2_ideal_DFT} for an illustration.

\subsection{Offset} \label{sec:2_offset}

Offsets in interleaved ADCs originate both from device mismatches in the sub-ADCs themselves (e.g., comparator offsets), and from offsets among the parallel track-and-hold stages used at large interleaving factors.
As offsets are additive, we can model the output of the interleaved ADC as
\begin{align}
    y^\textnormal{o}(t) = \textstyle y^\textnormal{id}(t) + \sum_k o_{k \modulo N}\delta \!\left( t - \frac{k}{\fs} \right)\!,
\end{align}
where $y^\textnormal{id}(t)$ is the ideally-sampled output and $o_n$ the offset of the \nth sub-ADC.
Applying the FT, we obtain
\begin{align} \label{eq:2_offset}
    \textstyle Y^\textnormal{o}(f) = Y^\textnormal{id}(f) +  \fs \sum_k \tilde{o}_{k \modulo N} \delta\!\left(f - k\frac{\fs}{N}\right)\!,
\end{align}
with $\tilde{\bmo}$ the DFT of the offset sequence $\bmo$.
We conclude that offsets among the sub-ADCs lead to spurs at multiples of $\fs/N$ whose amplitudes correspond to the DFT of the offset sequence, as illustrated in~\fref{fig:2_offset}.
Due to their additive nature, offset-induced spurs are independent of the input signal and constrain the dynamic range of the system.

\subsection{Gain Mismatch} \label{sec:2_gain}

Another common non-ideality in interleaved ADCs are gain mismatches, which 
arise from sub-ADC gain errors (e.g., capacitor ratio mismatches in SAR ADCs) and from mismatches in device or routing losses preceding the sub-ADCs.
We model the output of the interleaved ADC as
\begin{align}
    y^\textnormal{g}(t) &= \textstyle x(t) \sum_k (1+g_{k \modulo N})\delta \!\left( t - \frac{k}{\fs} \right) \nonumber\\
        &=\textstyle y^\textnormal{id}(t) + x(t)  \sum_k g_{k \modulo N}\delta \!\left( t - \frac{k}{\fs} \right)\!,
\end{align}
where $g_n$ is the relative gain mismatch of the \nth sub-ADC.
The FT of the output is therefore
\begin{align} \label{eq:2_gain}
    \textstyle Y^\textnormal{g}(f) = Y^\textnormal{id}(f) 
                        + X(f) \ast \fs \sum_k \tilde{g}_{k \modulo N} \delta\!\left(f - k\frac{\fs}{N}\right)\!,
\end{align}
with $\tilde{\bmg}$ the DFT of the mismatch sequence $\bmg$.

Similar to~\fref{eq:2_offset}, the result in \fref{eq:2_gain} features a Dirac comb with amplitudes corresponding to the DFT of the mismatch sequence.
However, the comb is now \emph{convoluted} to the FT of the input signal, leading to residual aliases from the $\fs/N$ rate, as illustrated in~\fref{fig:2_gain}.
For a single-tone input at $f_\textnormal{in}$, those replicas are spurs located at $n\fs/N\pm f_\textnormal{in}$ with $n$ an integer.
In practice, the DC component of $\tilde{\bmg}$, i.e., the average gain error, can be neglected, as it has little impact on the main signal component and does not significantly affect the spectrum.

\subsection{Timing Skew} \label{sec:2_skew}

Timing skew causes each sub-ADC to sample at a small time offset from its nominal sampling instant, so the output depends on the continuous-time behavior of the input.
In time domain, the output of the \nth sub-ADC with a skew $s_n$~is
\begin{align} \label{eq:2_skew_time}
    y^\textnormal{s}_n(t) =& \textstyle x(t-s_n) \sum_k \delta\!\left(t-k\frac{N}{\fs}-\frac{n}{\fs}\right)\!.
\end{align}

Assuming that the input $x(t)$ is band-limited to $\fmax$, and that the skew of each sub-ADC is small compared to the period $1/\fmax$, the FT of $x(t-s_n)$ can be approximated as
\begin{align}
    \textstyle X(f)\, \eexp^{-2\pi jfs_n} \approx X(f)(1-2\pi jfs_n).
\end{align}
The FT of~\fref{eq:2_skew_time} can then be approximated as 
\begin{align}
     \textstyle Y^\textnormal{s}(f) \!\approx\!  Y^\textnormal{id}(f) \! -\! \fs (2\pi jf X(f)) \!\ast\! 
            \sum_k \tilde{s}_{k \modulo N} \delta\!\left(f \!-\! k\frac{\fs}{N}\right)\!,
\end{align}
where $\tilde{\bms}$ the DFT of the timing skew sequence $\bms$.

Analogous to the case of gain mismatches, aliases appear with amplitudes determined by the DFT of the skew sequence, but in this case, a differentiated version of the input is replicated (cf.~\fref{fig:2_skew}).
The effect of timing skew thus grows linearly with frequency: doubling the input frequency for a given skew sequence increases the alias power by $6\,$dB.
By Bernstein’s inequality~\cite{Pinsky_2009,Zygmund_2003}, the band-limited signal with the largest derivative under an amplitude constraint is a single tone at~$\fmax$, therefore representing the worst-case input for skew analysis.


\section{Spur-Level Statistics} \label{sec:3}

In~\fref{sec:2}, we have shown that the impacts of offset, gain, or skew mismatches are proportional to the DFT of the mismatch sequences.
Since mismatches in an interleaved ADC are fixed after production (or drift slowly), they act as device-specific constants rather than random noise.
As a result, average metrics across realizations are not informative for design.
To link a production yield target to a mismatch variance, one instead needs the cumulative density function (CDF) of the power of each spur or replica as a function of the mismatch variance.

\begin{figure}[tp] 
	\centering
	\includegraphics[width=0.4\textwidth]{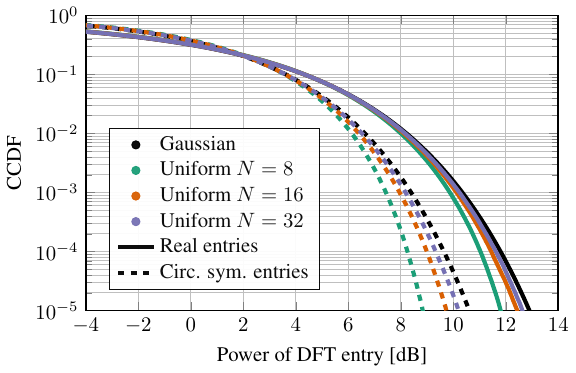}
	\vspace{-0.3cm}    
	\caption{Complementary CDF (CCDF) of the squared magnitude of DFT entries for i.i.d.~uniform sequences (various $N$) normalized to unit variance of the magnitude, compared with the Gaussian case. Already at $N=16$, the Gaussian approximation is accurate within $1$\,dB at $10^{-4}$ probability.}
	\label{fig:3_unif}
\end{figure}

We will assume that the mismatch sequences are i.i.d.~zero-mean real Gaussian random variables and derive the distribution of the DFT of such a vector.

\begin{lem} \label{lem:1}
    Let $\bmx$ be a vector of even length $N$ containing realizations of i.i.d. zero-mean real-valued Gaussian random variables, each of variance~$\sigma^2$. Then, the entries of its DFT \mbox{$\tilde{\bmx}=\bF\bmx$} are distributed as follows:
    \begin{align} \label{eq:3_lemma}
    \tilde{x}_k \sim
    \left\{
        \begin{array}{ll}
            \textstyle \normal\!\left(0, \frac{\sigma^2}{N}\right) \: & k=0,N/2 \\
            \textstyle \jpg\!\left(0, \frac{\sigma^2}{N}\right) \: & k=1,\dots, N/2-1,
        \end{array}\right.
    \end{align}
    where the entries in~\fref{eq:3_lemma} are mutually independent, and ${\tilde{x}_k = \tilde{x}_{N-k}^*}$ for $k=N/2+1,\ldots,N-1$.\footnote{If $N$ is odd, then the result is similar but without a Nyquist bin: the DC bin is a real-valued Gaussian and the rest of the DFT entries are complex-valued circularly-symmetric Gaussians arranged in complex conjugate pairs.}
\end{lem}

\fref{lem:1} can be proven by noticing that, by linearity of the DFT, $\tilde{\bmx}$ is a Gaussian random vector, hence fully characterized by its covariance matrix $\bK$ and pseudo-covariance matrix $\bJ$~\cite{Schreier_2010}
\begin{align}
    \bK &= \textstyle \Ex{\tilde{\bmx}\tilde{\bmx}^H} 
                    = \bF\Ex{\bmx\bmx^H}\bF^H = \frac{\sigma^2}{N}\bI_N  ,   \label{eq:3_K}\\
    \bJ &=  \textstyle \Ex{\tilde{\bmx}\tilde{\bmx}^T}
                    = \sigma^2\bF^2 = \frac{\sigma^2}{N}\bP.       \label{eq:3_J}
\end{align}
Here, $\bP$ is the permutation matrix with ones at entries $(k,\ell)$ whenever $(k+\ell)\modulo N$, including the DC bin and, if $N$ is even, also the Nyquist bin on the diagonal.
The structure of $\bK$ and~$\bJ$ then allows us to conclude on the distribution of $\tilde{\bmx}$.

\subsection{CDF of Offset Spurs}

As shown in~\fref{sec:2_offset}, offsets among sub-ADCs lead to input-independent spurs at multiples of $\fs/N$.
We assume the offset sequence $\bmo$ to contain i.i.d.\ zero-mean Gaussians with variance $\sigma_{\!\bmo}^2$ per sub-ADC, and derive the CDF of the squared magnitude of the spurs to link a target yield to mismatch variance.
In order to ensure that our results can be applied directly by designers, we consider a single-sided spectrum: the power of the spurs originating from circularly-symmetric Gaussian distributions therefore include a factor of two. 
We express the spur power relative to the full-scale power as defined in~\fref{sec:1_notation}; another factor of two is accordingly applied to spur powers.

For real zero-mean Gaussians--corresponding to the spur at DC and, when $N$ is even, also the spur at the Nyquist frequency--the squared magnitude $p$ of the spur follows a chi-squared distribution with CDF
\begin{align} \label{eq:3_cdf_off_real}
    \textstyle F^\textnormal{o}_\textnormal{real}(p) = \erf\!\left(\sqrt{\frac{Np}{4\sigma_{\!\bmo}^2}}\right)    \qquad p\geq0,
\end{align}
where $10\log(p)$ is the spur power in $\mathrm{dBFS}$.
For the rest of the spurs, i.e., the circularly-symmetric Gaussian-distributed bins, their power $p$ follows an exponential distribution of CDF
\begin{align} \label{eq:3_cdf_off_circ}
    \textstyle F^\textnormal{o}_\textnormal{circ}(p) = 1-\eexp^{-\tfrac{Np}{4\sigma_{\!\bmo}^2}}    \qquad p\geq0.
\end{align}
As can already be observed in~\fref{eq:3_lemma}, doubling the interleaving factor $N$ statistically reduces the power of each spur by a factor of two.
However, as the number of spurs also doubles, the total spur power does not change.

\subsection{CDF of Gain-Mismatch Replicas}

As shown in~\fref{sec:2_gain}, gain mismatch among sub-ADCs leads to aliasing replicas.
We are therefore interested in the power of those replicas relative to the power of the input, i.e., in $\mathrm{dBc}$.
We assume the gain-mismatch sequence to contain i.i.d.\ zero-mean Gaussians with variance $\sigma_{\!\bmg}^2$.
Unlike the case of offset, no power scaling is required: the squared magnitudes of the DFT entries of the mismatch sequence directly yield the power ratios in $\mathrm{dBc}$.

The impact of the DC bin of the DFT of $\bmg$ can be ignored, but, in case $N$ is even, the Nyquist bin creates a replica. Analog to~\fref{eq:3_cdf_off_real}, we obtain
\begin{align} \label{eq:3_cdf_gain_real}
    \textstyle F^\textnormal{g}_\textnormal{real}(p) = \erf\!\left(\sqrt{\frac{Np}{2\sigma_{\!\bmg}^2}}\right)   \qquad p\geq0,
\end{align}
where $10\log(p)$ is the replica power in $\mathrm{dBc}$.
For replicas originating from circularly-symmetric Gaussians, we obtain
\begin{align} \label{eq:3_cdf_gain_circ}
    \textstyle F^\textnormal{g}_\textnormal{circ}(p) = 1-\eexp^{-\tfrac{Np}{\sigma_{\!\bmg}^2}}   \qquad p\geq0.
\end{align}

\subsection{CDF of Skew Replicas}

We have shown in~\fref{sec:2_skew} that skews have a very similar effect compared to gain mismatches.
If we consider the power of the \emph{already-differentiated} replicas,~\fref{eq:3_cdf_gain_real} and~\fref{eq:3_cdf_gain_circ} can directly be used replacing $\sigma_{\!\bmg}^2$ by $\sigma_{\!\bms}^2$, the skew variance of a sub-ADC.

Alternatively, we consider the scenario of a single-tone at $\fsig$.
In this case, we simply rewrite~\fref{eq:3_cdf_gain_real} and~\fref{eq:3_cdf_gain_circ} as
\begin{align} \label{eq:3_cdf_skew_real}
    \textstyle F^\textnormal{s}_\textnormal{real}(p) &=  \textstyle
            \erf\!\left(\sqrt{\frac{Np}{8\pi^2\fsig^2\sigma_{\!\bms}^2}}\right)   \qquad &p\geq0, \\
    \textstyle F^\textnormal{s}_\textnormal{circ}(p) &=\textstyle 1-\eexp^{-\tfrac{Np}{4\pi^2\fsig^2\sigma_{\!\bms}^2}}   & p\geq0,
\end{align}
where $10\log(p)$ is the replica power in $\mathrm{dBc}$.

\subsection{Combined CDF for Maximum Spur Constraints}

For many practical cases, specifications are given as total spurious-free dynamic range (SFDR), i.e., we are interested in the probability of the strongest spur being below a target level in terms of $\mathrm{dBFS}$ or $\mathrm{dBc}$.
Since all contributions are assumed to be independent (\fref{lem:1}), the combined CDF is the product of all contributions.

\begin{figure}
    \includegraphics[width=0.82\linewidth]{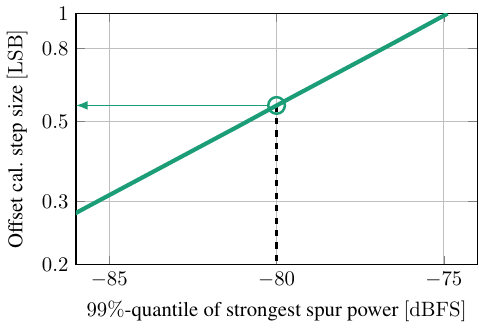}
    \vspace{-0.3cm}
    \caption{Offset calibration step size that ensures the strongest offset spur remains below a specified power limit with $99\%$ probability in a $12$-bit $16\times$ interleaved ADC. Note that $\textnormal{LSB}=2^{1-B}$ for a $B$-bit ADC.}
    \label{fig:4_spur_offs}
\end{figure}

Taking offset spurs as an example, we have two real Gaussian contributions (DC and $\fs/2$ spurs) and $N/2-1$ circularly-symmetric Gaussian contributions (for even $N$).
The combined CDF, i.e., the likelihood that the power of all those spurs remain under $10\log(p)\dBFS$ is then
\begin{align} \label{eq:3_offset_tot}
    \textstyle F^\textnormal{o}_\textnormal{tot}(p) &= 
        \textstyle F^\textnormal{o}_\textnormal{real}(p)^2  \cdot
        \textstyle F^\textnormal{o}_\textnormal{circ}(p)^{N/2-1} .
\end{align}
In many practical cases, spurs at DC and $\fs/2$ are not relevant, in which case the component $\textstyle F^\textnormal{o}_\textnormal{real}(p)^2$ in~\fref{eq:3_offset_tot} can be omitted.

The same reasoning can be applied for gain and timing skew spurs, but with only one real Gaussian contribution, since the DC (average) term is not relevant, leading to\footnote{As illustrated in~\fref{fig:2_gain}, we obtain $N\!-\!2$ replicas from the circularly-symmetric Gaussian entries of the DFT, but, due to their conjugate symmetric pairing, only half of them are independent contributors.}
\begin{align} \label{eq:3_gain_tot}
    \textstyle F^\textnormal{g}_\textnormal{tot}(p) &= 
        \textstyle F^\textnormal{g}_\textnormal{real}(p)  \cdot
        \textstyle F^\textnormal{g}_\textnormal{circ}(p)^{N/2-1}, \\ 
        \label{eq:3_skew_tot}
    \textstyle F^\textnormal{s}_\textnormal{tot}(p) &= 
        \textstyle F^\textnormal{s}_\textnormal{real}(p)  \cdot
        \textstyle F^\textnormal{s}_\textnormal{circ}(p)^{N/2-1}.
\end{align}
We note that we can freely consider specific combinations of spurs or replicas by including the relevant terms.

\subsection{Case of a Uniform Distribution} \label{sec:3_unif}

To mitigate the impact of spurs and replicas, interleaved ADCs are typically calibrated for sub-ADC mismatches.
Residual mismatches would then ideally follow i.i.d.\ uniform distributions of support $[-\Delta_\textnormal{cal}/2,\Delta_\textnormal{cal}/2]$, with $\Delta_\textnormal{cal}$ the calibration step size.
By the central limit theorem, the DFT of a mismatch sequence with i.i.d. entries approaches the Gaussian distribution of~\fref{lem:1} as $N$ grows.

\fref{fig:3_unif} provides Monte--Carlo results showing the inverse CDF of the squared magnitude of the DFT entries of i.i.d. uniform sequences, normalized to unit bin variance, compared with the Gaussian analytical results from~\fref{lem:1}.
We observe errors below $2$\dB, $1$\dB, and $0.5$\dB at $10^{-4}$ probability for $N=8$, $16$, and $32$, respectively.
We also note that the Gaussian approximation serves as a worst-case analysis: intuitively, this is because realizations of a Gaussian random variable are unbounded.
In most practical cases, the mismatch distribution is not exactly uniform due to noise in the estimates, resulting in an effective CDF in-between uniform and Gaussian.


\section{A Practical Example}

\begin{figure}
    \centering
    \includegraphics[width=0.95\linewidth]{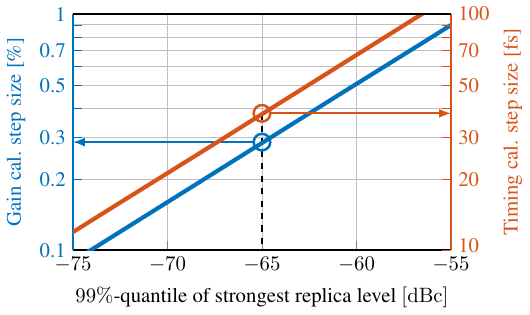}
    \vspace{-0.3cm}
    \caption{Gain and sampling instant calibration step size that ensures the strongest replica remains below a specified power limit relative to the input signal with $99\%$ probability in a $12$-bit $16\times$ interleaved ADC.}
    \label{fig:4_spur_combi}
\end{figure}

As an example, we consider a $12$-bit interleaved ADC for RF-sampling applications with $N\!=\!16$ sub-ADCs; our design goal is to determine the calibration step size~$\Delta_\textnormal{cal}$ required for each mismatch type.
Targets are set to $-80\dBFS$ for offset spurs and $-65\dBc$ for both gain and timing-skew spurs, for input frequencies up to $12\GHz$.
Offset spurs at DC and $f_s/2$ are excluded, while gain and skew replicas include the $\fs/2-\fsig$ component.
As discussed in~\fref{sec:3_unif}, we assume each mismatch to be Gaussian with variance~$\sigma_\textnormal{mis}^2=\Delta_\textnormal{cal}^2/12$.

Using \fref{eq:3_offset_tot}, \fref{eq:3_gain_tot}, and \fref{eq:3_skew_tot}, the $99\,\%$-quantile of the strongest spur or replica is obtained numerically as a function of~$\Delta_\textnormal{cal}$.
Results in~\fref{fig:4_spur_offs} indicate that a calibration step size of about half an LSB is required to meet the $-80\dBFS$ offset-spur limit.
As shown in~\fref{fig:4_spur_combi}, meeting the $-65\dBc$ target for gain and skew requires step sizes of approximately $0.27\,\%$ and $35\,\mathrm{fs}$, respectively, highlighting the stringent calibration accuracy needed for high-speed interleaved ADCs.


\section{Conclusions}

We have derived compact expressions describing the impact of offset, gain, and timing skew mismatch in interleaved ADCs.
We have further characterized the distribution of the power of spurs and replicas in the ADC output spectrum originating from those non-idealities under a Gaussian assumption; we have also shown that this Gaussian assumption is relevant in scenarios where mismatch calibration is employed, i.e., when mismatches follow a uniform distribution.
To illustrate the usefulness and flexibility of our results, we have derived a calibration step size requirement for a practical example.

\balance

\bibliographystyle{IEEEtran}
\bibliography{bib/IEEEfull, bib/confs-jrnls, bib/publishers, bib/bibliography}

\begin{thebibliography}{1}
\providecommand{\url}[1]{#1}
\csname url@samestyle\endcsname
\providecommand{\newblock}{\relax}
\providecommand{\bibinfo}[2]{#2}
\providecommand{\BIBentrySTDinterwordspacing}{\spaceskip=0pt\relax}
\providecommand{\BIBentryALTinterwordstretchfactor}{4}
\providecommand{\BIBentryALTinterwordspacing}{\spaceskip=\fontdimen2\font plus
\BIBentryALTinterwordstretchfactor\fontdimen3\font minus
  \fontdimen4\font\relax}
\providecommand{\BIBforeignlanguage}[2]{{%
\expandafter\ifx\csname l@#1\endcsname\relax
\typeout{** WARNING: IEEEtran.bst: No hyphenation pattern has been}%
\typeout{** loaded for the language `#1'. Using the pattern for}%
\typeout{** the default language instead.}%
\else
\language=\csname l@#1\endcsname
\fi
#2}}
\providecommand{\BIBdecl}{\relax}
\BIBdecl

\bibitem{Black_1980}
W.~Black and D.~Hodges, ``Time interleaved converter arrays,'' \emph{{IEEE}
  Journal of Solid-State Circuits}, vol.~15, no.~6, pp. 1022--1029, December
  1980.

\bibitem{adc_survey}
B.~Murmann, ``{ADC Performance Survey 1997-2025},'' [Online]. Available:
  \url{https://github.com/bmurmann/ADC-survey}.

\bibitem{Vogel_2005}
C.~Vogel, ``The impact of combined channel mismatch effects in time-interleaved
  {ADCs},'' \emph{{IEEE} Transactions on Instrumentation and Measurement},
  vol.~54, no.~1, pp. 415--427, February 2005.

\bibitem{Ghosh_2021}
S.~Ghosh and B.~D. Sahoo, ``Closed-form expression for the combined effect of
  offset, gain, timing, and bandwidth mismatch in time-interleaved {ADCs} using
  generalized sampling,'' \emph{{IEEE} Transactions on Instrumentation and
  Measurement}, vol.~70, pp. 1--12, October 2021.

\bibitem{Pinsky_2009}
M.~A. Pinsky, \emph{Introduction to Fourier analysis and wavelets}, ser.
  Graduate Studies in Mathematics.\hskip 1em plus 0.5em minus 0.4em\relax
  American Mathematical Society, 2009, vol. 102, pp. 122--124.

\bibitem{Zygmund_2003}
A.~Zygmund, \emph{Trigonometric Series}, ser. Cambridge Mathematical
  Library.\hskip 1em plus 0.5em minus 0.4em\relax Cambridge Univ. Press, 2003,
  p. 276.

\bibitem{Schreier_2010}
P.~J. Schreier, \emph{Statistical Signal Processing of Complex-Valued Data: The
  Theory of Improper and Noncircular Signals}.\hskip 1em plus 0.5em minus
  0.4em\relax Cambridge Univ. Press, 2010, pp. 30--58.

\end{thebibliography}

\balance

\end{document}